\documentclass[11pt,a4paper]{article}
\usepackage[draft]{minted}
\usepackage[hyperref]{acl2021}
\usepackage{times}
\usepackage{latexsym}

\usepackage{cite}
\usepackage{float}
\usepackage[utf8]{inputenc} 
\usepackage[T1]{fontenc}    
\usepackage{hyperref}       
\usepackage{url}            
\usepackage{booktabs}       
\usepackage{amsfonts}       
\usepackage{nicefrac}       
\usepackage{microtype}      
\usepackage{tcolorbox}
\usepackage{adjustbox}
\usepackage{multirow}
\usepackage{siunitx}
\setminted{fontsize=\tiny}
\definecolor{bg}{HTML}{282828}
\usepackage{pmboxdraw}
\usepackage{algorithmic}
\usepackage{listings}
\usepackage{graphicx}
\usepackage[linesnumbered,ruled,vlined]{algorithm2e}
\def\BibTeX{{\rm B\kern-.05em{\sc i\kern-.025em b}\kern-.08em
    T\kern-.1667em\lower.7ex\hbox{E}\kern-.125emX}}

\hypersetup{%
  colorlinks=true, 
  pdfborderstyle={/S/U/W 1}
}

\usepackage{paralist}

\begin{document}

\newcommand{\ie}{\textit{i.e.,}~}
\newcommand{\eg}{\textit{e.g.,}~}
\newcommand{\etc}{\textit{etc.}~}
\newcommand{\etal}{\textit{et al.}~}

\newcommand{\nb}[2]{
    \fbox{\bfseries\sffamily\scriptsize#1}
    {\sf\small$\blacktriangleright$\textit{#2}$\blacktriangleleft$}
}

\newcommand\MICHELE[1]{\textcolor{blue}{\nb{MICHELE}{#1}}}
\newcommand\DAWN[1]{\textcolor{green}{\nb{DAWN}{#1}}}
\newcommand\ALEXEY[1]{\textcolor{red}{\nb{ALEXEY}{#1}}}

\newcommand{\approach}{{\sc AthenaTest}\xspace}
\newcommand{\dataset}{{\sc Methods2Test}\xspace}

\newcommand\mycommfont[1]{\footnotesize\ttfamily\textcolor{blue}{#1}}
\SetCommentSty{mycommfont}

\SetKwInput{KwInput}{Input}                
\SetKwInput{KwOutput}{Output}              

\lstdefinestyle{myJavaStyle}{
  frame=tb,
  float=*,
  language=java,
  aboveskip=3mm,
  belowskip=3mm,
  showstringspaces=false,
  columns=flexible,
  basicstyle={\small\ttfamily},
  numbers=none,
  numberstyle=\tiny\color{gray},
  keywordstyle=\color{blue},
  commentstyle=\color{dkgreen},
  stringstyle=\color{mauve},
  frame=single,
  breaklines=true,
  breakatwhitespace=true,
  tabsize=3,
}

\title{Generating Code with the Help of Retrieved Template Functions and Stack Overflow Answers}
\author{
  Dawn Drain\thanks{~~Corresponding author}\\
  Microsoft Cloud and AI\\
  dawn.drain@microsoft.com\And
  
  Changran Hu\thanks{~~Work done as an intern at Microsoft} \\
  Microsoft Cloud and AI\\
  changran\_hu@berkeley.edu\And
  
  Chen Wu\\
  Microsoft Cloud and AI \\
  wu.chen@microsoft.com\AND
  
  Mikhail Breslav\\
  Microsoft Cloud and AI \\
  mikhail.breslav@microsoft.com\And
  
  Neel Sundaresan \\
  Microsoft Cloud and AI \\
  neels@microsoft.com
}
\maketitle


\begin{abstract}
We approach the important challenge of code autocompletion as an open-domain task, in which a sequence-to-sequence code generator model is enhanced with the ability to attend to reference code snippets supplied by a semantic code search engine.
In this work, we present a novel framework to precisely retrieve template functions as well as intent-snippet pairs and effectively train such a retrieval-guided code generator.
To demonstrate the effectiveness of our model designs, we perform extensive experiments with CodeSearchNet~\citep{CSN} which contains template functions and CoNaLa~\citep{conala} which contains Stack Overflow intent-snippet pairs.
We also investigate different retrieval models, including Elasticsearch, DPR~\citep{DPR}, and our fusion representation search model, which currently holds the number one spot on the CodeSearchNet leaderboard.
We observe improvements by leveraging multiple database elements and further gain from retrieving diverse data points by using Maximal Marginal Relevance.
Overall, we see a 4\% improvement to cross-entropy loss, a 15\% improvement to edit distance, and a 44\% improvement to BLEU score when retrieving template functions.
We see subtler improvements of 2\%, 11\%, and 6\% respectively when retrieving Stack Overflow intent-snippet pairs.
We also create a novel Stack Overflow-Function Alignment dataset, which consists of 150K tuples of functions and Stack Overflow intent-snippet pairs that are of help in writing the associated function, of which 1.7K are manually curated.

\end{abstract}

\section{Introduction}


Code autocompletion is an important and ubiquitous tool for enhancing developer productivity. Recent advances in NLP, especially the usage of large, pretrained sequence-to-sequence neural models~\citep{BART,T5,bi2020palm,song2019mass} have enabled code completion systems to write larger and more complicated code snippets, such as unit tests, bug-fixes, transcompiler and arbitrary methods \citep{unit_test_gen, pymt5, codexglue}. 

These models typically have an impoverished context compared to what professional developers can access.
At the most extreme, the input may only be a function's signature. In contrast, a developer may consult pre-existing implementations of related functions, look up API usages, and peruse solutions to common questions on sites like Stack Overflow. That is, writing code is in general an open-domain task, whereas researchers typically treat it as closed-book.










\begin{figure}[h]
\vspace{-0.2cm}
    \centering
\footnotesize
  Example of Retrieving a Template \\
\begin{minipage}[t]{0.45\textwidth}
\begin{minted}[escapeinside=||]{python}
# Input
def add_toc_to_epub(tocless_epub_path, write_path):
    """
    reads the epub file, adds a table of contents, 
    and saves it
    """
# Retrieved Template
def create_epub(epub_path):
    """
    Creates an example epub
    including metadata, toc, and css
    and saves it to epub_path 
    """

    book = epub.EpubBook()
    book.set_title('Sample book')

    # create chapter
    c1 = epub.EpubHtml(title='Intro', 
                file_name='chap_01.xhtml', lang='hr')
    c1.content=u'''<h1>Intro heading</h1>
                <p>Zaba je skocila u baru.</p>'''

    # add chapter
    book.add_item(c1)

    # define Table Of Contents
    book.toc = (epub.Link('chap_01.xhtml', 
                            'Introduction', 'intro'),
                 (epub.Section('Simple book'),
                 (c1, ))
                )

    # add default NCX and Nav file
    book.add_item(epub.EpubNcx())
    book.add_item(epub.EpubNav())

    # basic spine
    book.spine = ['nav', c1]

    # write to the file
    epub.write_epub('test.epub', book, {})
\end{minted}
\end{minipage} \\

\begin{minipage}[t]{0.45\textwidth} 
\begin{minted}[escapeinside=||]{python}
    
def add_toc_to_epub(tocless_epub_path, write_path):
    """
    reads the epub file, adds a table of contents, 
    and saves it
    """
    # Generated Code
    book = epub.read(epub(tocless_epub_path))
    book.toc = tuple(book.get_items_of_type(
                                epub.EpubHtml))
    book.spine = ['nav'] + book.spine
    epub.write_epub(write_path, book, {})
\end{minted}
\caption{We consider the task of generating a function's body given its signature and docstring. Given that input, we use the docstring as a query to retrieve a helpful reference function or code snippet, create\_epub in this manually created example, which we concatenate with the input and then feed to our generator model.}
\end{minipage} \\
\vspace{-0.4cm}
\label{fig:template_example}
\end{figure}

\begin{figure}
    \centering
    \includegraphics[width=0.5\textwidth]{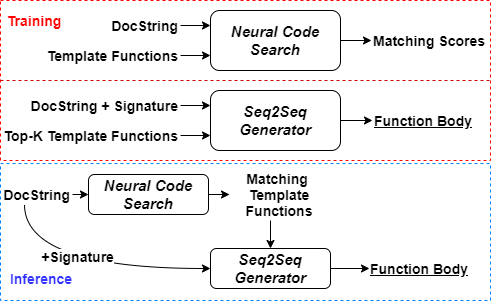}
    \caption{Flow charts of training and inference based on retrieving template functions. A flexible framework for retrieval-guided code generation which consists of a neural code search model for template function retrieval and a template guided code generator. The training of the template functions searching and the code snippet generator is decoupled, and they work cooperatively under the help of the retrieval system.}
    \label{fig:framework}
\end{figure}

To address the problem, we similarly consider the task of writing functions given a signature and docstring, and attempt to bridge this gap by enabling our models to retrieve template functions from GitHub and answers to relevant questions from Stack Overflow. For both of these databases, we experiment with several different retrieval models.
In all cases we use a pre-trained sequence-to-sequence transformer as the generator, and condition generation on the target function's signature and docstring as well as the retrieved documents. We attain our best results when using our fusion representation model to retrieve as many template functions from CodeSearchNet \citep{CSN} as fit in the context window, with the templates re-ranked according to their Maximal Marginal Relevance scores \citep{MMR}.
Figure~\ref{fig:framework} depicts the training and inference procedures of our framework.

Rather than retrieve from all of Stack Overflow, our work uses CoNaLa as the Stack Overflow database \citep{conala}. CoNaLa consists of intent-snippet pairs, where each intent is the title of a Stack Overflow question concerning Python, and each code snippet is extracted from the answering question using a neural model.
More specifically, we follow~\citep{conala_filtering}, use the most confidently extracted 100K intent-snippet pairs, including 2.3K that were manually curated. 

Experiments show that our method significantly improves the informativeness of the generated code as well as their relevance to the corresponding query.
When retrieving template functions, we see a 4\% improvement to cross-entropy loss, a 15\% improvement to edit distance, and a 44\% improvement to BLEU score. We see subtler improvements of 2\%, 11\%, and 6\% respectively when retrieving Stack Overflow answers.
In addition, We conduct extensive ablation studies to evaluate the improvement from different retrieval model design.
To summarize, our contributions are as follows:
\begin{itemize}
    \item We propose a flexible two-stage framework for retrieval-guided code generation. 
    \item We propose a semantic code search model for template functions retrieval.
    \item We propose a template-aware code generator that can handle template functions.
    \item We create a novel Stack Overflow-Function Alignment dataset, which consists of 150K tuples of functions and Stack Overflow intent-snippet pairs that are of help in writing the associated function, of which 1.7K are manually curated. We will release the dataset for future research.
\end{itemize}

\section{Related Work}
Some prior work has considered code generation through the lens of search. Similarly to our approach, \citep{Liang} experiment with retrieving a template method given the input function signature and docstring, resulting in a large boost in BLEU score from 19.2 to 34.7. In order to train the retriever, they attempt to minimize the edit distance between the retrieved template and the ground truth function. In the CodeSearchNet challenge, \citep{CSN} focus more specifically on curating a database of functions and on performing code-search using natural language queries. Additionally, they host a leaderboard for various retrieval models, including several baselines they have open-sourced. They take a function's docstring as a query, excluding parameter annotations, and form a database out of functions denuded of their docstrings. In order to train the retriever, they give a binary reward, with one point for retrieving the function the docstring belongs to, and zero points otherwise.

Several works have also explored using Stack Overflow as a database. As SO questions and answers can be very long, a standard approach is to restrict to the question's title and a code block extracted from the corresponding answer. Among these works, both CODENN \citep{codenn} and StaQC \citep{staqc} match the question title to the entire code block, while CoNaLa \citep{conala} matches the question title more precisely to the key lines in the code block. Our work uses CoNaLa as the Stack Overflow database to retrieve, as it allows for a higher information density, and enables retrieving more intent-snippet pairs.

In NLP more broadly, much work has focused on open-domain question answering with Wikipedia as a database. 
Typically the answer is at most a few words long, and the retriever is rewarded for returning an article containing that exact substring. One state of the art retriever is DPR \citep{DPR}, which is based on BERT \citep{bert}. DPR outperforms a strong Lucene-BM25 baseline by 9-15\%, and attains superior results with as few as one thousand training examples. 
The follow-up work RAG pipelines DPR with BART as a generator, similarly to our approach \citep{RAG}. 
Open-domain scores compare quite favorably to those of models constrained to closed-book question answering. For example, the 11B-parameter T5 achieves a score of 34.5 EM on Natural Questions and the 770M-parameter T5 achieves a score of 28.9 EM \citep{T5}. Meanwhile, RAG achieves a score of 44.5 EM with only 626M parameters \citep{RAG}.




\section{Retrieval-guided Code Generation}

\subsection{Task Definition}
The problem is as follows: Given a function's signature and docstring as input, produce the rest of the function's body. We treat this as an open-domain problem, and experiment both with retrieving template functions and relevant answers from Stack Overflow. For simplicity we focus on Python functions, although our approach is language agnostic.

Consider a database $D = \{d_1,\dots,d_m\}$, consisting of either template functions or Stack Overflow answers, and a set of target functions $T = \{t_1,\dots,t_n\}$ that begin with the signatures and docstrings $\{s_1,\dots,s_n\}$. 
In the closed-book setting we train a code generator $G_c$ so as to minimize the cross-entropy:
$$\mathbb{E}_T \text{log}P(t_i | s_i,G_c).$$ In our open domain setting, we seek a retriever model $R$ and generator $G_o$ to minimize:
$$\mathbb{E}_T \text{log}P(t_i | s_i,G_o,d_{i,1},\dots,d_{i,k_i}),$$
where $\{d_{i,1},\dots,d_{i,k_i}\}$ is the set of documents retrieved by $R$ given $s_i$. 

In practice we condition retrieval solely on the docstring, although we condition on both the signature and docstring during generation along with the retrieved documents.


\subsection{Retrieval Models}\label{sec:Retrieval Models}

All of our neural retrieval models are based around nearest neighbor lookup between a vectorized query and a database of vectorized function templates or Stack Overflow answers. We take the function's docstring as the query, as that is the same input used in CodeSearchNet. In all cases we use separate encoders for the queries and database elements, although in general one could reuse a single encoder. 

Rather than perform exact nearest neighbor lookup, in practice we use approximate nearest neighbor search using ANNOY \citep{annoy}. ANNOY works by partitioning the entire vector space into subspaces through constant random projections and then performing nearest neighbor search in the lower-dimensional subspaces.
In our experiments we use angular distance as the distance metric, set $n\_trees = 1000$, and set $search\_k = 10000$. 

In the following section, we will first introduce three retrieval models on CodeSearchNet in \ref{sec:es_baseline}, \ref{sec:nbow_baseline} and \ref{sec:fusion_representation} respectively, and then presents a retrieval model on Stack Overflow in \ref{sec:dpr}.
\subsubsection{ElasticSearch Baseline}
\label{sec:es_baseline}
We use Elasticsearch as our baseline retrieval model \citep{gormley2015elasticsearch}. ElasticSearch is a widely used, inverted index-based search engine. Here we use the NDCG evaluation results reported by CodeSearchNet.

\subsubsection{NBoW Bi-encoder Baseline}
\label{sec:nbow_baseline}
We use the Joint Vector Representations for Code Search proposed by CodeSearchNet as another baseline. Specifically, we use their open-sourced NBoW \citep{CSN} model as a bi-encoder.

The model consists of two NBoW encoders, one for transforming the natural language documentation of a function into a vector representation, and one for transforming the corresponding code into a vector representation. The natural language and code are then mapped into a single vector space by training.

Before being put into the model, the data was first processed by splitting the code by identifiers in the code and splitting the natural language documentation using BPE. Secondly, the tokenized sequences are entered into the corresponding encoder separately to obtain a series of vector representations. After that, the series of vector representations are condensed into a single vector through one of mean-pooling, max-pooling, and an attention-like weighted sum mechanism. Finally, the dot product between the vectors obtained from the natural language documentation and the code is computed. To train the model, natural language and code snippets from within the same function are rewarded, while discrepant pairs are penalized.

\subsubsection{Bi-encoder with Fusion Representation}
\label{sec:fusion_representation}

We present our fusion representation model, which is currently ranked number one on the CodeSearchNet leaderboard. We build on CSN's open-sourced NBoW baseline, and more than double its mean-NDCG score while maintaining the dimensionality of the embedding space and only adding one linear alignment layer and two triplets of fusing weights to the parameter count.

Recall that the NBoW baseline trains two encoders, one for the docstring query and one for the template functions comprising the database. Each encoder condenses its hidden states $\mathrm{H}$ to a single vector $\mathbf{h}$ using one of mean-pooling, max-pooling, or a self-attention type of weighted average. We reuse those three types of pooling, with the modification that we standardize the third type of pooling by adding a softmax to the attention scores as below, where $w_h$ is a learnable weight vector.
$$ \gamma={softmax}\left(\mathbf{w}_{\mathrm{h}}^{\top} \cdot \mathrm{H}\right) $$
$$\mathbf{h}_{attn}=\sum_{\mathrm{j}} \gamma_{\mathrm{j}} \cdot \mathrm{H}_{: \mathrm{j}}, \forall \mathrm{j} \in[1, \ldots, \mathrm{m}]$$

We fuse all three vectors using a linear combination $\beta \cdot \sum_{k \in (mean,max,attn)} w_{k} \mathbf{h}_{k}$ with learnable weights $w_{mean}, w_{max}, \text{ and } w_{attn}$. $\beta$ allows the language encoder to scale the entire fused vector.
Finally, we add a learnable linear layer to the code-encoder only, to help align the code-embedding space with the docstring-embedding space. Empirically, the largest weight is given to $w_{attn}$.

\subsubsection{Dense Passage Retrieval}
\label{sec:dpr}

Dense Passage Retrieval (DPR) is a bi-encoder model proposed to solve the retrieval part of open domain question answering \citep{DPR}. As with our fusion representation model, DPR trains separate encoders to embed the queries and database documents. Both encoders of DPR are initialized as pretrained BERT models.

We fine-tune the open-sourced DPR model on our proposed dataset SOFA-auto-mined (See \hyperref[sec:SOFA]{SOFA} for more details), using each function's docstring as the query and the corresponding intent-snippet pair as the target database document. As DPR warmstarts using a BERT model pretrained on English, we form the intent-snippet embedding using only the intent question-title. We take all SO-functions pairs in SOFA-auto-mined as positive examples. For each method we randomly select 15 SO intent-snippet pairs as negative examples. 
Intuitively, training directly on retrieving intent-snippets given a function's docstring as input will lead to better downstream performance than training with question titles (the intents) as input.


\subsection{Generative Model}
We use BART-large as the body of our Seq2Seq architecture. 
BART-large is a performative transformer with 406M-parameters, including twelve encoder layers and twelve decoder layers \citep{BART}. We initialize the weights from the open-sourced BART-large which was pretrained on English corpus, and continue pretraining on a mix of masked Python code and Stack Overflow question-answer pairs. We additionally add a small number of randomly initialized vocabulary embeddings for common whitespace tokens such as the four-space, eight-space, and two-tab tokens, which increases the effective context window length and throughput.





\section{Data Preparation}\label{sec:Data}

\subsection{Pretraining Corpus}
Following best practice in NLP, we conduct extensive pretraining before finetuning on the task of method generation. We take advantage of the open-sourced model BART \citep{BART}, which was pretrained for 40 epochs on 160GB of English text using a denoising objective. In order to adapt the model to the domain of source code, we augment its tokenizer with whitespace tokens, such as the four-space token and the two-tab token. We then perform task-adaptive pretraining on a mix of 4GB of Stack Overflow question-answers pairs, corresponding to 1.35M questions tagged `python' with answers that have non-negative scores, and 27GB of raw python masked in the style of T5 \citep{T5}. We train for two weeks on sixteen 32GB Tesla V100 GPU's.


\par
\subsection{Code Generation Dataset}
We introduce the two datasets we use as databases to retrieve from and reference during generation.

\subsubsection{Template Functions Database}

We first adopt the CodeSearchNet \citep{CSN} dataset of python functions that have docstrings, and perform the following data cleaning.

First, we remove all class methods. Class methods are generally highly dependent on their fellow member functions, which are not included as part of CSN. Furthermore, we do not need to train a retrieval model to select this class information, as classes are already localized within a single file.

Second, we tokenize functions based on punctuation and spaces, and remove functions with more than 150 tokens, which are often too complex to generate in their entirety. Note that we use the more standard BPE tokens during training and inference.

We split this cleaned version of CSN into training, validation, and test sets, and use the training set to build the function template database.


\begin{table}[]
\centering 
\begin{tabular}{llll}
\hline
Dataset        & Train  & Validation & Test \\
Non-class CSN & 119480 & 1000       & 1000 \\
\hline \\

\end{tabular}
\caption{\label{tab:table-name}Number of CodeSearchNet functions after filtering. We exclude class-methods, methods longer than 150 tokens after splitting on punctuation and whitespace, and all CSN methods that lack docstrings.}
\end{table}

\subsubsection{Stack Overflow Database}

We performed preliminary experiments using all Stack Overflow Python questions as the database, and attained underwhelming results. Manual inspection revealed that very few answers were actually of relevance to the target function body we were trying to generate, even after cursory filtering based on the number of stars each answer received. 


We thus turned towards a more extensively filtered subset of Stack Overflow answers, as explored previously by \citep{conala}, in their work leading to the creation of the dataset CoNaLa. Similar to our goal, their aim was to retrieve code snippets from Stack Overflow that best answered a natural-language query. 

CoNaLa is a superior database for two reasons. First, rather than consisting of questions and answers in their entirety, CoNaLa keeps only the question title, which is typically sufficient to gather the question's intent, as well as the most relevant code snippet from the question's answer, which is determined using machine learning. Restricting to these intent-snippet pairs increases information density, and allows us to pack many more retrieved answers into the context window.

Second, CoNaLa is filtered to only contain 600K intent-snippet pairs in python that address ``how-to" questions, such as "how to flatten a list in python", in contrast to questions addressing a particular bug or error message, or open-ended questions discussing more philosophical topics. 

Following~\citep{conala_filtering}, we use the 100K highest rated intent-snippet pairs as our SO-database, along with 2.3K intent-snippet pairs that CoNaLa manually curated in order to train their relevance scoring model.

\begin{table*}[]
\resizebox{1.0\textwidth}{!}{\begin{minipage}{\textwidth}
\centering 
\begin{tabular}{ll}
\hline
\textbf{Intent}  & How to convert a list of multiple integers into a single integer? \\ \hline
\textbf{Snippet} & sum(d * 10 ** i for i, d in enumerate(x{[}::-1{]}))
             \\ \hline 
\textbf{Intent}  & Averaging the values in a dictionary based on the key \\ \hline
\textbf{Snippet} & [(i, sum(j) / len(j)) for i, j in list(d.items())]               \\ \hline \\

\end{tabular}
\caption{\label{tab:table-name} Example intent-snippet pairs from the CoNaLa dataset. Each intent is the title of a Stack Overflow question, and each snippet is a code snippet that answers the question, as determined using a neural model.}
\end{minipage}}
\end{table*}

\subsection{SOFA, a Novel Code Search Dataset}
\label{sec:SOFA}



The input to our retrieval models is a function’s docstring, which represents a different distribution from Stack Overflow question titles. Thus we are motivated to create a dataset that more directly caters to our intended workflow than does CoNaLa. Towards this aim we create a dataset consisting of (function, (intent, snippet)) pairs, where each SO intent-snippet is selected so as to be of help in generating the CSN method. We term this the Stack Overflow-Function Alignment (SOFA) dataset.

In order to pair an intent-snippet with a CSN method, we use our implementation of the similarity measure introduced in Aroma \citep{luan2019aroma}. We manually inspect the suggested method for each of the 2.3K intent-snippet pairs manually curated in forming CoNaLa and filter out 600 pairs for which the intent-snippet is unhelpful for generating the corresponding function or even itself trivial. To scale up we first rank the 600K intent-snippet pairs in CoNaLa by their confidence scores. We next keep the top 10K pairs, which we consider to be of high quality. Then for each intent-snippet pair we retrieve 15 snippets from CSN that are most similar, and discard snippets which are not in our filtered version of CSN.

\begin{figure}
    \centering     \includegraphics[width=0.5\textwidth]{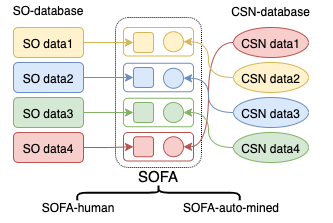}
    \caption{Overview of SOFA, which we use to train the retriever DPR-SO. We build on the database CoNaLa, which consists of intent-snippet pairs, where each intent is the title of a Stack Overflow question, and each code snippet is extracted from that question's answer. We automatically match functions from CodeSearchNet with relevant intent-snippet pairs using the code search tool Aroma. For the subset of CoNaLa intent-snippet pairs that were manually curated, we manually discard unhelpful matches provided by Aroma.}
    \label{fig:SOFA}
\end{figure}

\begin{table*}[]
\centering
\begin{tabular}{lllll}
\cline{1-2}
\multicolumn{1}{l}{\textbf{SO-intent}} & How to convert a list of multiple integers into a single integer? &  &  &  \\ \cline{1-2}
\multicolumn{1}{l}{\textbf{SO-snippet}} & sum(d * 10 ** i for i, d in enumerate(x{[}::-1{]})) &  &  &  \\ \cline{1-2}
\multicolumn{1}{l}{\textbf{CSN-function}} & \begin{tabular}[c]{@{}l@{}} 

def lower\_endian\_to\_number(l, base):\\         \quad  """Helper function: convert a list of digits in the given base to a number."""\\        \quad  return sum({[}d * (base**i) for i, d in enumerate(l){]})

\end{tabular} &  &  &  \\ \cline{1-2}
 &  &  &  & 
\end{tabular}
\caption{Example from our manually curated SOFA benchmark. We include this CoNaLa intent-snippet pair as it is considered helpful in generating the CodeSearchNet function.} 
\end{table*}


\begin{figure}[h]
\vspace{-0.2cm}
    \centering
\begin{adjustbox}{width=.49\textwidth}
\footnotesize
\begin{tabular}{cc}
\toprule
  Stack Overflow Retrieval Example \\
\midrule
\begin{minipage}[t]{0.45\textwidth}
\begin{minted}[escapeinside=||]{python}
# Input
def nii_ugzip(imfile, outpath=''):
    """
    Uncompress *.gz file
    """
# Retrieved SOFA intent-snippet pairs
# Loading a file into a numpy array
with pythonfile = open('file.txt', 'rb')

# Generate temporary file names 
# without creating actual file in Python
f = gzip.open(temp_file_name, 'wb')

# gzip a file in Python
f = gzip.open('/home/joe/file.txt.gz', 'wb')

# Using openpyxl to read file from memory
file = open('path/to/file.xlsx', 'rb')
wb = openpyxl.load_workbook(filename=file)

# gzip a file in Python
f_in = open('/home/joe/file.txt')
\end{minted}
\end{minipage}\\
\midrule
\begin{minipage}[t]{0.45\textwidth}
\begin{minted}[escapeinside=||]{python}
# Hypothesis without context:
imfile = os.path.abspath(imfile)
if not os.path.exists(outpath):
    os.makedirs(outpath)
with gzip.open(imfile, 'rb') as f:
    nii_data = f.read()
return nii_data
\end{minted}
\end{minipage}\\
\midrule
\begin{minipage}[t]{0.45\textwidth}
\begin{minted}[escapeinside=||]{python}
# Hypothesis with context:
if not os.path.exists(outpath):
    os.mkdir(outpath)
with gzip.open(imfile, 'rb') as infile:
    with gzip.open(outpath, 'wb') as outfile:
        outfile.write(infile.read())
\end{minted}
\caption{Example of the benefits of retrieving intent-snippet pairs from Stack Overflow. The generated hypothesis with context for this example learns to write into the outpath provided in the function signature, whereas the hypothesis without context merely reads the file and returns it.}
\end{minipage}\\
\end{tabular}


\end{adjustbox}
\vspace{-0.4cm}
\label{fig:first_not_found_fix}
\end{figure}

\section{Experiments and Results Analysis}
We experiment with manipulating four different variables:
\begin{enumerate}
    \item The retrieval model used (Elasticsearch, Bi-encoder with Fusion representation, or DPR)
    \item The number of function templates retrieved
    \item Retrieving the top-k results independently vs. using Maximal Marginal Relevance (MMR)
    \item The choice of database (CodeSearchNet or Stack Overflow)
\end{enumerate}

\subsection{Retrieval Models}

We experiment with three different retrieval models on CodeSearchNet: Elasticsearch, our own fusion representation model, and the DPR model \citep{DPR}. 

In Table \ref{table:fusionrepresentation} we compare our fusion representation model against Elasticsearch and NBoW on CodeSearchNet. Notably, Elasticsearch outperforms the neural baselines as reported in the CodeSearchNet paper. The fusion representation model outperforms the ElasticSearch baseline by 4.7\% on average NDCG across the six languages, and 6.6\% on python NDCG only.
The Fusion Representation model also outperforms the current state of the art NBoW model by 1.25\% on average NDCG, and 0.9\% on python NDCG only, which suggests that our proposd fusion strategy can better accommodate natural language and programming language to the same vector space.

\begin{table}[]
\centering
\begin{tabular}{llllll}
\cline{1-4}
\textbf{}                                                        & ES & NBoW   & Fusion Rep &  &  \\ \cline{1-4}
\begin{tabular}[c]{@{}l@{}}NDCG\\ (mean)\end{tabular}   & 33.7\%        & 37.2\% & \textbf{38.4\%}                &  &  \\
\begin{tabular}[c]{@{}l@{}}NDCG\\ (python)\end{tabular} & 40.6\%        & 46.3\% & \textbf{47.2\%}                &  &  \\ \cline{1-4}
\end{tabular}
\caption{Comparison between Elasticsearch, Neural bag-of-words retriever (NBoW), and our fusion representation model on the Codesearchnet benchmark. Note that we draw the NBoW results from the CSN leaderboard, rather than from the weaker NBoW model open-sourced as part of CSN. The mean NDCG score is taken across six different programming languages.}
\label{table:fusionrepresentation}
\end{table}

\subsection{Number of Examples Retrieved}
When experimenting with CodeSearchNet, we consider four different limits on the number of template functions retrieved.

No-Retrieve: As a baseline, we consider only using the query as the generator's input.

Single-Retrieve: Select the method with the highest similarity to the query and use its signature, docstring, and method body as contextual information.

Threshold-Retrieve: More strictly than in the single-retrieve condition, only present a template function if its retrieval score is greater than the median score across all data points. We hypothesize that foregoing mediocre retrievals will help prevent mode collapse, in which the generator learns to ignore the retrieved data.

Full-Retrieve: We retrieve as many template functions as fit in the generator's context window, presented in decreasing order of retrieval score. Since the generator is based on BART-large, its context window is 1024 tokens wide, some of which is consumed by the query function's signature and docstring. 

In table \ref{table:number_retrieved} we experiment with varying the number of template functions retrieved. We observe monotonically increasing performance with the number retrieved, except for our control condition in which we retrieve random templates and observe a small drop in performance. Filling out the entire context window with retrieved templates is roughly twice as effective as retrieving a single template. In the Threshold-Retrieve condition we also experiment with only retrieving a single template if its retrieval score passes a threshold, which does worse than simply always retrieving a template.

\begin{table}[]
\centering 
\resizebox{1.0\linewidth}{!}{
\begin{tabular}{lllll}
\hline

\textbf{}                                                       & \textbf{Ppl}  & \textbf{EDist} & \textbf{\begin{tabular}[c]{@{}l@{}}BLEU-4\\ (BPE)\end{tabular}} & \textbf{BLEU-4} 
\\ \hline
No retrieve                                                     & 1.25          & 1.46           & 17.40                                                           & 5.76                          
\\ \hline
Threshold retrieve                                                   & 1.25          & 1.43           & 18.61                                                           & 6.98                       
\\ \hline
Single retrieve                                                 & 1.24          & 1.35           & 19.83                                                           & 7.69                       
\\ \hline
Full retrieve                                                   & \textbf{1.21}          & \textbf{1.30}           & \textbf{22.17}                                                           & \textbf{9.66}                        
\\ \hline
\begin{tabular}[c]{@{}l@{}}Full random \\ retrieve\end{tabular} & 1.24          & 1.46           & 16.83                                                           & 5.52                       
\\ \hline \\
\end{tabular}
}
\caption{
Results for retrieving multiple templates. In all cases we use our fusion representation model as the retriever and CSN as the database. We observe monotonically increasing performance with the number of retrieved templates, and no benefit from only including documents whose retrieval scores surpass a threshold.}
\label{table:number_retrieved}
\end{table}

\subsection{Returning top-k independently vs. MMR}
There is a well-known diversity problem when retrieving multiple search results. That is, there is little benefit in retrieving a set of redundant templates or explanations, and we would achieve better results by retrieving a more comprehensive set of data points. One popular technique for retrieving diverse results is Maximal Marginal Relevance (MMR) \citep{MMR}. Given the query $q$, we first retrieve the closest datapoint $d_1$ as usual. Inductively, having already retrieved data points $\{d_1,\dots,d_i\} =: D_i \subset D$, we next retrieve
$$
d_{i+1}:= argmax_{d \in D\setminus D_i} \big(\lambda(d \cdot q)+(1-\lambda)\max_{d_i \in D_i} d \cdot d_i \big)
$$
where `$\cdot$' denotes the dot product. Setting $\lambda = 1$ recovers the standard $k$-nearest neighbor approach. For $\lambda < 1$ the second term penalizes documents that are similar to any that have already been retrieved. In our experiments we set $\lambda = .5$.


As a control, we also experiment with eschewing a retriever and instead collecting functions at random. As for greedy retrieval and MMR, we provide as many functions as fit in the 1024-token context window.

In table \ref{table:mmr} we achieve our strongest results by not merely greedily adding templates based on their retrieval scores, but instead by greedily adding successive templates according to their marginal relevance scores. When reranking based on MMR, we add a diversity term to each retrieval score based on the template's distance from previously retrieved examples. MMR doubles the benefit from retrieving more than one template, as measured by the edit distance and BLEU scores.

\begin{table}[]
\centering 
\resizebox{1.0\linewidth}{!}{
\begin{tabular}{lllll}
\hline
\textbf{}                                                       & \textbf{Ppl}  & \textbf{EDist} & \textbf{\begin{tabular}[c]{@{}l@{}}BLEU-4\\ (BPE)\end{tabular}} & \textbf{BLEU-4}
\\ \hline
Full retrieve                                                   & 1.21          & 1.30           & 22.17                                                           & 9.66                      
\\ \hline
MMR retrieve                                                     & \textbf{1.20} & \textbf{1.24}  & \textbf{25.09}                                                  & \textbf{11.87}             
\\ \hline
\end{tabular}
}
\caption{\label{tab:table-name}
Results for retrieving templates independently vs. using Maximal Marginal Relevance. In all cases we use our fusion representation model as the retriever and CSN as the database.  We achieve our best results when using MMR to incrementally add diverse documents, in contrast to the standard greedy approach.}
\label{table:mmr}
\end{table}

\section{Discussion}
\subsection{Choice of Database}
We experiment between retrieving relevant template functions from the CodeSearchNet dataset and retrieving relevant intent-snippet pairs from Stack Overflow. 
We achieve poor results when retrieving from all Stack Overflow questions tagged `python', and thus restrict to the refined subset shared by CoNaLa.  During validation we restrict to using the training set as the database so as to prevent leakage. During training we additionally exclude any retrieved template method that is identical to the target method, so as to prevent degenerate behavior on the part of the retriever.

When retrieving template functions from CodeSearchNet, we achieved our strongest results when retrieving as many templates as possible and reranking them using MMR. Thus we use the same approach when experimenting with Stack Overflow as a database. 
The intent-snippet pairs we retrieve give a relatively small boost, as reported in table \ref{table:compare-retrieve-SO-for-gen}. Indeed, we achieve better results using a single template from CodeSearchNet compared to filling out the entire context window with intent-snippet pairs taken from Stack Overflow.

Perhaps the Stack Overflow database gives comparatively worse results since the retriever is weaker. If so, we might speculate that this is because it is so much harder to construct ground truth pairings between functions and intent-snippets. Alternatively the differing results could be due to the quality of the databases themselves. Search is most useful when recalling details about the long tail of obscure API usages and when making small modifications to pre-existing code. We think this information is better supplied by functions found in the wild than by the kind of supplementary documentation found on Stack Overflow. 

\begin{table}[]
\centering 
\resizebox{1.0\linewidth}{!}{
\begin{tabular}{lllll}
\hline
\textbf{}    & \textbf{Ppl}  & \textbf{EDist} & \textbf{\begin{tabular}[c]{@{}l@{}}BLEU-4\\ (BPE)\end{tabular}} & \textbf{BLEU-4} 
\\ \hline
No retrieve  & 1.25          & 1.46           & 17.40                                                           & 5.76            
\\ \hline
Full ES   & 1.24          & 1.44           & 17.20                                                           & 5.82          
\\ \hline
Full fusion representation & \textbf{1.23} & \textbf{1.33}  & \textbf{18.54}                                                  & \textbf{6.32}   
\\ \hline
Full DPR-SO  & 1.24          & 1.36           & 17.56                                                           & 5.52          
\\ \hline \\
\end{tabular}
}
\caption{\label{table:model_comparison} Evaluation of Stack Overflow as a database. We retrieve as many answers as fit in the context window and filter those answers using MMR, inspired by the success of that approach when using template functions from CSN.}
\label{table:compare-retrieve-SO-for-gen}
\end{table}

\section{Conclusion}
In this paper, we presented a novel framework pipelining search with generation for retrieval-guided code generation. Our method uses a neural code search model for extracting template functions and intent-snippet pairs and uses a robust programming language generator for code generation. 
The two components are trained separately to allow more flexibility and efficiency. 
In order to integrate Stack Overflow intent-snippet pairs with the task of function generation we also created a novel code search dataset, SOFA, which we will release for future research.
Experiments show that our method significantly outperforms several strong baselines.

\bibliographystyle{acl_natbib}
\bibliography{main}

\begin{thebibliography}{20}
\expandafter\ifx\csname natexlab\endcsname\relax\def\natexlab#1{#1}\fi

\bibitem[{cod(2020)}]{codexglue}
 2020.
\newblock Codexglue: A benchmark dataset and open challenge for code
  intelligence.

\bibitem[{Bernhardsson(2020)}]{annoy}
Erik Bernhardsson. 2020.
\newblock \href {https://pypi.org/project/annoy/} {\emph{Annoy: Approximate
  Nearest Neighbors in C++/Python}}.
\newblock Python package version 1.16.0.

\bibitem[{Bi et~al.(2020)Bi, Li, Wu, Yan, and Wang}]{bi2020palm}
Bin Bi, Chenliang Li, Chen Wu, Ming Yan, and Wei Wang. 2020.
\newblock Palm: Pre-training an autoencoding\&autoregressive language model for
  context-conditioned generation.
\newblock \emph{arXiv preprint arXiv:2004.07159}.

\bibitem[{Carbonell and Goldstein(1998)}]{MMR}
Jaime Carbonell and Jade Goldstein. 1998.
\newblock The use of mmr, diversity-based reranking for reordering documents
  and producing summaries.
\newblock In \emph{Proceedings of the 21st annual international ACM SIGIR
  conference on Research and development in information retrieval}, pages
  335--336.

\bibitem[{Clement et~al.(2020)Clement, Drain, Timcheck, Svyatkovskiy, and
  Sundaresan}]{pymt5}
Colin~B. Clement, Dawn Drain, Jonathan Timcheck, Alexey Svyatkovskiy, and Neel
  Sundaresan. 2020.
\newblock \href {https://doi.org/10.18653/v1/2020.emnlp-main.728} {Pymt5:
  multi-mode translation of natural language and python code with
  transformers}.
\newblock In \emph{Proceedings of the 2020 Conference on Empirical Methods in
  Natural Language Processing, {EMNLP} 2020, Online, November 16-20, 2020},
  pages 9052--9065. Association for Computational Linguistics.

\bibitem[{Devlin et~al.(2019)Devlin, Chang, Lee, and Toutanova}]{bert}
Jacob Devlin, Ming-Wei Chang, Kenton Lee, and Kristina Toutanova. 2019.
\newblock \href {http://arxiv.org/abs/1810.04805} {Bert: Pre-training of deep
  bidirectional transformers for language understanding}.

\bibitem[{Gormley and Tong(2015)}]{gormley2015elasticsearch}
Clinton Gormley and Zachary Tong. 2015.
\newblock \emph{Elasticsearch: the definitive guide: a distributed real-time
  search and analytics engine}.
\newblock " O'Reilly Media, Inc.".

\bibitem[{Hashimoto et~al.(2018)Hashimoto, Guu, Oren, and Liang}]{Liang}
Tatsunori~B Hashimoto, Kelvin Guu, Yonatan Oren, and Percy~S Liang. 2018.
\newblock A retrieve-and-edit framework for predicting structured outputs.
\newblock In \emph{Advances in Neural Information Processing Systems}, pages
  10052--10062.

\bibitem[{Husain et~al.(2019)Husain, Wu, Gazit, Allamanis, and
  Brockschmidt}]{CSN}
Hamel Husain, Ho-Hsiang Wu, Tiferet Gazit, Miltiadis Allamanis, and Marc
  Brockschmidt. 2019.
\newblock Codesearchnet challenge: Evaluating the state of semantic code
  search.
\newblock \emph{arXiv preprint arXiv:1909.09436}.

\bibitem[{Iyer et~al.(2016)Iyer, Konstas, Cheung, and Zettlemoyer}]{codenn}
Srinivasan Iyer, Ioannis Konstas, Alvin Cheung, and Luke Zettlemoyer. 2016.
\newblock \href {https://doi.org/10.18653/v1/P16-1195} {Summarizing source code
  using a neural attention model}.
\newblock In \emph{Proceedings of the 54th Annual Meeting of the Association
  for Computational Linguistics (Volume 1: Long Papers)}, pages 2073--2083,
  Berlin, Germany. Association for Computational Linguistics.

\bibitem[{Karpukhin et~al.(2020)Karpukhin, O{\u{g}}uz, Min, Wu, Edunov, Chen,
  and Yih}]{DPR}
Vladimir Karpukhin, Barlas O{\u{g}}uz, Sewon Min, Ledell Wu, Sergey Edunov,
  Danqi Chen, and Wen-tau Yih. 2020.
\newblock Dense passage retrieval for open-domain question answering.
\newblock \emph{arXiv preprint arXiv:2004.04906}.

\bibitem[{Lewis et~al.(2019)Lewis, Liu, Goyal, Ghazvininejad, Mohamed, Levy,
  Stoyanov, and Zettlemoyer}]{BART}
Mike Lewis, Yinhan Liu, Naman Goyal, Marjan Ghazvininejad, Abdelrahman Mohamed,
  Omer Levy, Ves Stoyanov, and Luke Zettlemoyer. 2019.
\newblock \href {http://arxiv.org/abs/arXiv:1910.13461} {Bart: Denoising
  sequence-to-sequence pre-training for natural language generation,
  translation, and comprehension}.

\bibitem[{Lewis et~al.(2020)Lewis, Perez, Piktus, Petroni, Karpukhin, Goyal,
  K{\"u}ttler, Lewis, Yih, Rockt{\"a}schel et~al.}]{RAG}
Patrick Lewis, Ethan Perez, Aleksandara Piktus, Fabio Petroni, Vladimir
  Karpukhin, Naman Goyal, Heinrich K{\"u}ttler, Mike Lewis, Wen-tau Yih, Tim
  Rockt{\"a}schel, et~al. 2020.
\newblock Retrieval-augmented generation for knowledge-intensive nlp tasks.
\newblock \emph{arXiv preprint arXiv:2005.11401}.

\bibitem[{Luan et~al.(2019)Luan, Yang, Barnaby, Sen, and
  Chandra}]{luan2019aroma}
Sifei Luan, Di~Yang, Celeste Barnaby, Koushik Sen, and Satish Chandra. 2019.
\newblock Aroma: Code recommendation via structural code search.
\newblock \emph{Proceedings of the ACM on Programming Languages},
  3(OOPSLA):1--28.

\bibitem[{Raffel et~al.(2019)Raffel, Shazeer, Roberts, Lee, Narang, Matena,
  Zhou, Li, and Liu}]{T5}
Colin Raffel, Noam Shazeer, Adam Roberts, Katherine Lee, Sharan Narang, Michael
  Matena, Yanqi Zhou, Wei Li, and Peter~J. Liu. 2019.
\newblock \href {http://arxiv.org/abs/arXiv:1910.10683} {Exploring the limits
  of transfer learning with a unified text-to-text transformer}.

\bibitem[{Song et~al.(2019)Song, Tan, Qin, Lu, and Liu}]{song2019mass}
Kaitao Song, Xu~Tan, Tao Qin, Jianfeng Lu, and Tie-Yan Liu. 2019.
\newblock Mass: Masked sequence to sequence pre-training for language
  generation.
\newblock \emph{arXiv preprint arXiv:1905.02450}.

\bibitem[{Tufano et~al.(2020)Tufano, Drain, Svyatkovskiy, Deng, and
  Sundaresan}]{unit_test_gen}
Michele Tufano, Dawn Drain, Alexey Svyatkovskiy, Shao~Kun Deng, and Neel
  Sundaresan. 2020.
\newblock Unit test case generation with transformers.
\newblock \emph{arXiv preprint arXiv:2009.05617}.

\bibitem[{Xu et~al.(2020)Xu, Jiang, Yin, Vasilescu, and
  Neubig}]{conala_filtering}
Frank~F. Xu, Zhengbao Jiang, Pengcheng Yin, Bogdan Vasilescu, and Graham
  Neubig. 2020.
\newblock \href {https://doi.org/10.18653/v1/2020.acl-main.538} {Incorporating
  external knowledge through pre-training for natural language to code
  generation}.
\newblock In \emph{Proceedings of the 58th Annual Meeting of the Association
  for Computational Linguistics}, pages 6045--6052, Online. Association for
  Computational Linguistics.

\bibitem[{Yao et~al.(2018)Yao, Weld, Chen, and Sun}]{staqc}
Ziyu Yao, Daniel~S. Weld, Wei{-}Peng Chen, and Huan Sun. 2018.
\newblock \href {http://arxiv.org/abs/1803.09371} {Staqc: {A} systematically
  mined question-code dataset from stack overflow}.
\newblock \emph{CoRR}, abs/1803.09371.

\bibitem[{Yin et~al.(2018)Yin, Deng, Chen, Vasilescu, and Neubig}]{conala}
Pengcheng Yin, Bowen Deng, Edgar Chen, Bogdan Vasilescu, and Graham Neubig.
  2018.
\newblock Learning to mine aligned code and natural language pairs from stack
  overflow.
\newblock In \emph{2018 IEEE/ACM 15th International Conference on Mining
  Software Repositories (MSR)}, pages 476--486. IEEE.

\end{thebibliography}

\end{document}